\documentclass[preprint]{elsarticle}
\usepackage{latexsym, graphicx}

\begin{document}

\begin{frontmatter}

\title{Notes on nonlocal projective measurements in relativistic systems}
\author{Shih-Yuin Lin}
\ead{sylin@cc.ncue.edu.tw}
\address{Department of Physics, National Changhua University of Education,Changhua 50007, Taiwan}
\date{December 2, 2013}

\begin{abstract}
In quantum mechanical bipartite systems, naive extensions of von Neumann's projective measurement to nonlocal variables can produce superluminal signals and thus violate causality. We analyze the projective quantum nondemolition state-verification in a two-spin system and see how the projection introduces nonlocality without entanglement. For the ideal measurements of ``R-nonlocal" variables, we argue that causality violation can be resolved by introducing further restrictions on the post-measurement states, which makes the measurement ``Q-nonlocal". After we generalize these ideas to quantum mechanical harmonic oscillators, we look into the projective measurements of the particle number of a single mode or a wave-packet of a relativistic quantum field in Minkowski space. It turns out that the causality-violating terms in the expectation values of the local operators, generated either by the ideal measurement of the ``R-nonlocal" variable or the quantum nondemolition verification of a Fock state, are all suppressed by the IR and UV cutoffs of the theory. Thus relativistic quantum field theories with such projective measurements are effectively causal.
\end{abstract}



\end{frontmatter}


\section{Introduction}

Following von Neumann's prescription, a projective measurement of a variable will collapse or project a quantum state
to one of the variable's eigenstates instantaneously and simultaneously in the whole time-slice that the wave-function is defined on. 
Such wave-function collapse seems to be incompatible with relativity, where the transmission of physical information 
cannot be faster than light and the simultaneity of the events depends on the reference frame. 
However, once a variable is defined locally in space, the projective measurements of it will respect causality. 
Even in relativistic quantum field theory (RQFT), projective measurements of isolated, spatially localized variables and 
the corresponding wave-functional collapse will be consistent with relativity \cite{Lin12}.

When extended to the variables nonlocal in space, the situation becomes more complicated. 
It has been shown that projective measurements of nonlocal variables can produce superluminal signals and so violate causality
both in quantum mechanics (QM) with finite speed of physical information \cite{AAV86, PV94, BP99}
and in RQFT \cite{So93, BGKP02}.
Aharonov, Albert, and Vaidman thus introduced the causality principle, which identifies an operator $\hat{O}$
associated with some variable in a theory 
as measurable only if, for every local observable defined at some space point ${\bf x}$, the probability of any specific outcome of 
measuring that local observable (or its expectation value) obtained at some moment $t$ after the measurement of the operator $\hat{O}$ 
is independent of any local operation before the measurement of $\hat{O}$ and spacelike separated from $(t,{\bf x})$ \cite{AAV86, PV94}.
They further showed that, by using two entangled pointlike probes or detectors each interacting locally in spacetime with 
one element of a QM atom-pair separated in space, some nonlocal variables of the two-atom system 
can be measured instantaneously on the probes without knowing the details of each atom. 
Here a duration of the probe-atom interaction and a traveling time for the probes from the interaction region 
to the future measurement event
are necessary, so the whole process is causal \cite{AAV86, PV94, BP99}. 
Similar methods have been successfully applied to more complicated QM systems \cite{GV01, Va03}.

It is certainly interesting to extend these methods with local interactions and measurements
to describe the measurements of nonlocal variables in RQFT. Our attempts in this direction applying the Unruh-DeWitt detector theory
will be presented elsewhere \cite{Lin14}. In this notes, however, we are going in another direction: To see 
how far we can go in describing the measurements of nonlocal variables in QM and RQFT using von Neumann's prescription of the 
projective measurement, though the realization of such a nonlocal measurement is an outstanding issue. 

The paper is organized as follows. 
We review an example given by Sorkin \cite{So93} in Section 2. In Section 3 we analyze the 
quantum nondemolition state-verification measurement in a two-spin system to see how it violates causality. Then we
look into the ideal measurements of nonlocal variables in the same system in Section 4.
In Section 5 we extend our discussion to harmonic oscillators, then we look into the nonlocal measurement in RQFT in Section 6.
With the new knowledge we learned there, we revisit Sorkin's example in Section 7.
Finally we give a summary of our findings in Section 8. 
For the readers not familiar with RQFT in the Schr\"odinger picture, 
we offer some details about the wave-functional of a scalar field in Appendix A.

Before we start our discussion, let us distinguish two different nonlocalities.
In \cite{AAV86} a quantum measurement is said to be nonlocal if it extracted only part of the information in a composite QM system,
e.g., measuring the total spin angular momentum of two particles, 
without knowing the exact states of each element of the system. 
The effect of such kind of measurement is limited or localized in a subspace of the Hilbert space of the system, while literally
the degrees of freedom being measured may or may not be separated in the position space. To be more precise, 
below we call an operation or a measurement of a variable ``R-(non)local" (R for relativistic) if the variable measured is defined
(non)locally in the position space, and ``Q-(non)local" (Q for quantum) if the effect of the measurement is (non)localized 
in a subspace of the Hilbert space of the composite system.
For example, we say the measurement of the $z$-component of the total spin angular momentum $\vec{s}^{}_A +\vec{s}^{}_B$ of two spins 
$\vec{s}^{}_A$ and $\vec{s}^{}_B$ located at the same point 
is R-local and Q-local, since the subspace of $\vec{s}^{}_A-\vec{s}^{}_B$ in the Hilbert space is not affected by the measurement. 
Later we will see that if the two spins are separated in the position space, the R-nonlocal measurement may be Q-local or Q-nonlocal,
depending on the prescription.

\section{An R-nonlocal measurement in RQFT}
\label{SorkinEx}

In Ref. \cite{So93}, Sorkin demonstrated that a projective measurement of an one-particle state 
in RQFT can violate causality. In his example,
Sorkin introduced the ``yes" projector $|1_{\bf p}\rangle \langle 1_{\bf p}|$ and the ``no" projector 
${\bf 1}-|1_{\bf p}\rangle \langle 1_{\bf p}|$  for a one-particle state of some field mode with the wave vector ${\bf p}$, then 
argued that a local operation on a field degree of freedom $\phi(\bf{x})$ at some position ${\bf x}$ in space 
right before the ``yes-no" measurement on some time-slice can affect the expectation value of another field degree of freedom 
$\phi({\bf y})$ at a distance right after the measurement. Such an influence is almost instantaneous
and could be used to send superluminal signal, hence such kind of quantum measurement is not allowed
in formulating RQFT.
\begin{figure}[h]
\center{\includegraphics[width=8cm]{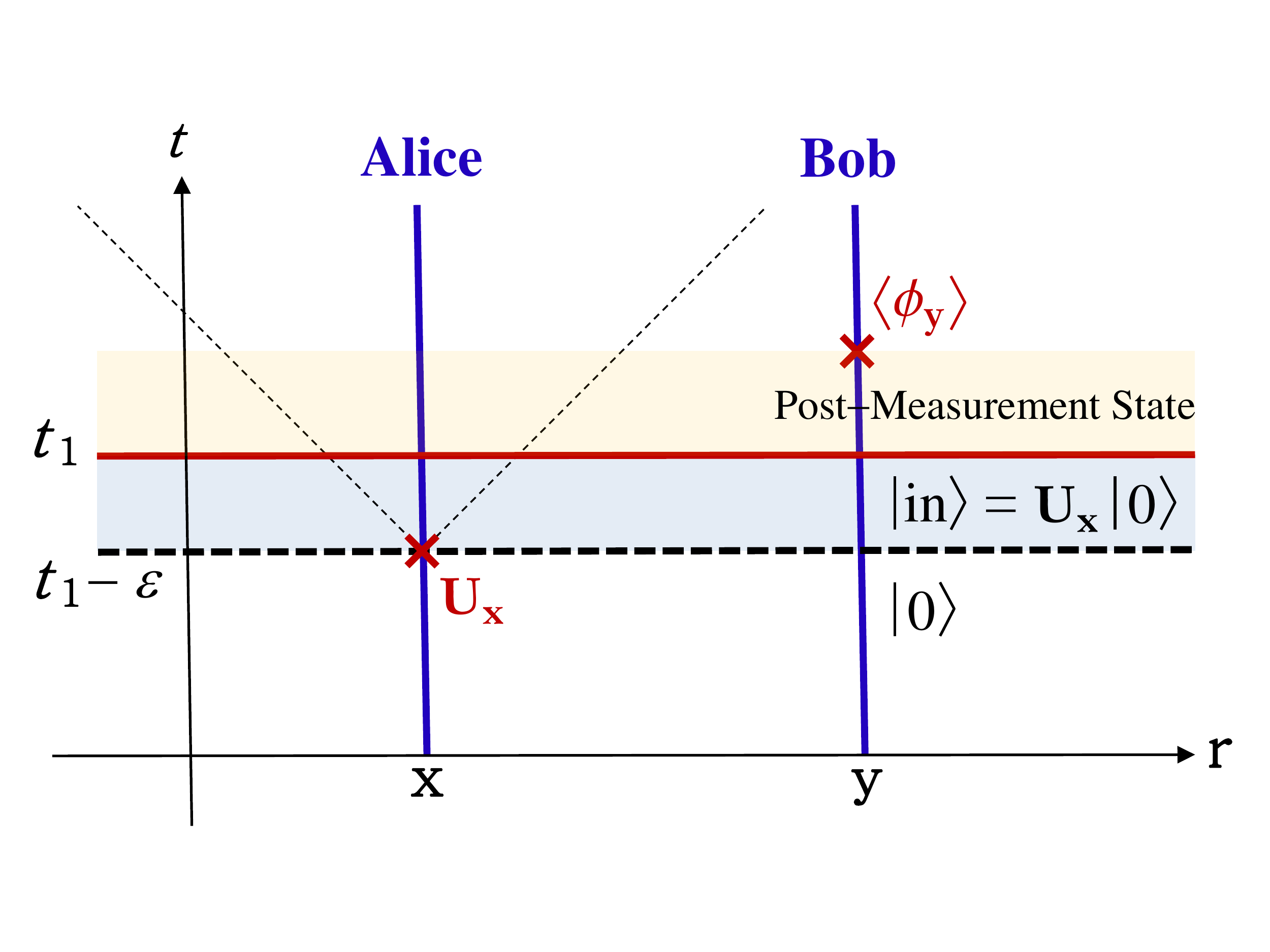}}
\caption{The setup. A wavefunction(al) collapse occurs at $t=t_1$ by measuring a nonlocal variable. 
Here $\varepsilon \to 0+$ and the event that Bob obtains $\langle \phi_{\bf y} \rangle$ is 
outside of the future lightcone of the event that Alice performs the operation ${\bf U}_{\bf x}$.}
\label{setup}
\end{figure}

More explicitly, consider the setup in Fig.\ref{setup}. 
Suppose the initial state of the field is the Minkowski vacuum $|0\rangle$. 
At $t_1-\varepsilon$,  $\varepsilon \to 0+$, Alice at ${\bf x}$ applies an R-local operation 
\begin{equation}
  {\bf U}_{\bf x} = e^{i\lambda\phi_{\bf x}/\hbar} \label{Ux}
\end{equation} 
to the field, so that the field state right before the measurement 
(the pre-measurement state, pre-MS) becomes $|M_-\rangle = {\bf U}_{\bf x}|0\rangle$.
At $t_1$, a measurement for checking whether the field is in the one-particle state of mode ${\bf p}$, $| 1_{\bf p} \rangle$,  
is performed: If the outcome is ``yes", the field state will collapse to the post-measurement state (post-MS) $| 1_{\bf p} \rangle$
(``nondemolition"); if ``no", the post-MS will become $({\bf 1}-|1_{\bf p}\rangle \langle 1_{\bf p}|)|M_-\rangle$ up to a normalization factor, 
which is orthogonal to $| 1_{\bf p} \rangle$. Then right after $t_1$, Bob at some point ${\bf y}\not={\bf x}$ will find 
\begin{eqnarray}
  \langle \phi_{\bf y} \rangle &=& 
  P_{\rm yes} {\langle M_-| 1_{\bf p} \rangle \langle 1_{\bf p}  | \phi_{\bf y}| 1_{\bf p}  \rangle \langle 1_{\bf p}  
  |M_- \rangle \over \langle M_- | 1_{\bf p} \rangle \langle 1_{\bf p}  | M_- \rangle   }+ \nonumber\\ & &
  P_{\rm no} {\langle M_- |\left( {\bf 1} - | 1^{}_{\bf p}  \rangle \langle 1^{}_{\bf p}  | \right) 
    \phi_{\bf y}\left({\bf 1} - | 1^{}_{\bf p} \rangle \langle 1^{}_{\bf p}  |\right)| M_- \rangle \over
    \langle M_- |\left( {\bf 1} - | 1^{}_{\bf p}  \rangle \langle 1^{}_{\bf p}  | \right) 
    \left({\bf 1} - | 1^{}_{\bf p} \rangle \langle 1^{}_{\bf p}  |\right)| M_-\rangle } \nonumber\\ 
  &=& \langle M_- | 1_{\bf p} \rangle \langle 1_{\bf p}  | \phi_{\bf y}| 1_{\bf p}  \rangle \langle 1_{\bf p}  | M_- \rangle + 
  \nonumber\\ & & \langle M_- |\left( {\bf 1} - | 1^{}_{\bf p}  \rangle \langle 1^{}_{\bf p}  | \right) 
  \phi_{\bf y}\left({\bf 1} - | 1^{}_{\bf p} \rangle \langle 1^{}_{\bf p}  |\right)| M_- \rangle \nonumber\\
  &=& -2 {\rm Re}\, \left\{  \langle M_-  | \phi_{\bf y} | 1_{\bf p} \rangle \langle 1_{\bf p}| M_- \rangle \right\}
\label{result1}
\end{eqnarray}
since the probabilities of finding the results ``yes" and ``no" are $P_{\rm yes} = \left|\langle 1_{\bf p}  | M_- \rangle\right|^2$ 
and $P_{\rm no} = \left|\left({\bf 1} - | 1^{}_{\bf p} \rangle \langle 1^{}_{\bf p}  |\right) | M_- \rangle\right|^2 = 1-P_{\rm yes}$,
while $\langle M_- | \phi_{\bf y}| M_- \rangle =0$ and $\langle 1_{\bf p} | \phi_{\bf y}| 1_{\bf p} \rangle =0$.
It can be shown that $d\langle \phi_{\bf y} \rangle/d\lambda \not= 0$ as $\lambda\to 0$ \cite{So93}, 
so $\langle \phi_{\bf y} \rangle$ depends on $\lambda$ in general, no matter how far Bob and Alice are apart.
This suggests that Alice could manipulate the values of $\lambda$ to send superluminal signal to Bob,
if such an R-nonlocal (but Q-local) ``yes-no" measurement were allowed 
\footnote{Note that, to obtain the expectation value $\langle \phi_{\bf y} \rangle$, one has to have many copies of the quantum states of the field.
Here we simply assume that this is possible.}.

In von Neumann's prescription, when a system is measured by an external agency with an operator corresponding to a variable,
the quantum state of the system will collapse instantaneously and discontinuously 
to one of the eigenstates associated with the operator. 
We say a process is an {\it ideal measurement} if it can be described in this way.
When the eigenstates of that operator are non-degenerate, namely, the post-MS of the corresponding 
ideal measurement form a complete orthogonal set, we call it a {\it complete orthogonal measurement} \cite{BGNP01}.
If there are degenerate eigenstates indistinguishable by the operator, the above prescription will not fix
a specific complete set of the eigenstates as the post-MS. 

The above measurement in Sorkin's example is an ideal measurement (with eigenstates $| 1^{}_{\bf p} \rangle$ and 
$({\bf 1}-|1_{\bf p}\rangle \langle 1_{\bf p}|)|M_-\rangle$ \cite{So93}, but not a complete orthogonal measurement.
Such a ``yes-no" measurement is called the quantum nondemolition state-verification (QNDSV) in Refs. \cite{AAV86, PV94}, 
where it has been shown that this kind of measurements 
can produce superluminal signals even in simple QM systems with two localized objects separated in the position space. 
Below we look into an example to see how this happens. 

\section{QNDSV on two spins}
\label{QNDSV2spin}

Consider two spin-particles A and B held by Alice and Bob, respectively, separated at a distance.
Similar to the setup in Fig. \ref{setup}, suppose before $t_1-\varepsilon$ ($\varepsilon \to 0+$) the two-spin system is in the initial state
\begin{equation}
  |\psi\rangle = |\uparrow^{}_A \uparrow^{}_B\rangle 
\end{equation}
where $\sigma_{j}^z |\uparrow_j\rangle = |\uparrow_j\rangle$ and $\sigma_{j}^z |\downarrow_j\rangle = -|\downarrow_j\rangle$
with the Pauli matrices $\vec{\sigma}_j$ and $j=A,B$. 
At $t_1-\varepsilon$, Alice can either choose to do nothing on the state $|\psi\rangle$, or choose to rotate the spin of particle A
by a unitary transformation ${\bf U}^{}_A$ so that
\begin{equation}
  |\psi'\rangle = {\bf U}^{}_A |\psi\rangle = |\rightarrow\hspace{-1.5mm}^{}_A \uparrow^{}_B\rangle 
\end{equation}
where $\sigma_{j}^x |\rightarrow\hspace{-1.5mm}^{}_j\rangle = |\rightarrow\hspace{-1.5mm}^{}_j \rangle$ and $\sigma_{j}^x 
|\leftarrow_j\rangle = -|\leftarrow_j\rangle$.
Suppose at $t_1$, a QNDSV measurement is performed to verify whether the state is in $|\uparrow^{}_A \rightarrow\hspace{-1.5mm}^{}_B\rangle$,
which is separable. Then after the QNDSV measurement, the expectation value of $\hat{s}^z_B$ for particle B becomes
\begin{eqnarray}
  \langle \hat{s}_B^z\rangle &=& {\rm Tr}
     \left( {\hbar\over 2}\sigma_B^z|\uparrow^{}_A \rightarrow\hspace{-1.5mm}^{}_B\rangle \langle\uparrow^{}_A \rightarrow\hspace{-1.5mm}^{}_B|\rho
    |\uparrow^{}_A \rightarrow\hspace{-1.5mm}^{}_B\rangle \langle\uparrow^{}_A \rightarrow\hspace{-1.5mm}^{}_B| \right) + \nonumber\\
  & & {\rm Tr}\left[ {\hbar\over 2} \sigma_B^z\left({\bf 1}-|\uparrow^{}_A \rightarrow\hspace{-1.5mm}^{}_B\rangle \langle\uparrow^{}_A
    \rightarrow\hspace{-1.5mm}^{}_B|\right) \rho \left({\bf 1}-|\uparrow^{}_A \rightarrow\hspace{-1.5mm}^{}_B
    \rangle \langle\uparrow^{}_A \rightarrow\hspace{-1.5mm}^{}_B|\right)\right] \nonumber\\
  &=& \left\{ \begin{array}{lcl}
    0 & {\rm for} & \rho = |\psi\rangle \langle \psi|, \\
    \hbar/4 & {\rm for} & \rho = |\psi'\rangle \langle \psi'|.
  \end{array} \right. \label{ExpSigBz}
\end{eqnarray}
So $|\psi\rangle$ and $|\psi'\rangle$ under different local operations on particle A by Alice
will give different expectation values $\langle \sigma_B^z\rangle$ of particle B at Bob's place 
right after the QNDSV, even if $|\psi\rangle$ and $|\psi'\rangle$ are both separable.
If such a measurement were allowed by nature, using an ensemble of $|\psi\rangle$ 
Alice would be able to send superluminal binary signals to Bob 
by encoding her information in the local operations ${\bf 1}^{}_A$ and ${\bf U}^{}_A$. 

Where is the superluminal signaling coming into play in this example with all the states involved separable? 
If there is no measurement ever performed at $t_1$, then Bob will find the expectation value of $\sigma_B^z$
equal to the one taken after a complete orthogonal measurement with the post-MS 
\begin{equation}
  \left\{ | +^{}_A \uparrow^{}_B\rangle, |+^{}_A \downarrow^{}_B\rangle, 
  |-^{}_A \uparrow^{}_B\rangle, |-^{}_A \downarrow^{}_B\rangle \right\},
\end{equation}
where $|\pm^{}_A\rangle$ are eigenstates of $\vec{\sigma}^{}_A$ in arbitrary direction with eigenvalues $\pm 1$, 
no matter what Alice has done on particle A.
But now the above QNDSV measurement at $t_1$ enforces $|\psi\rangle$ to collapse to either
$|\uparrow^{}_A \rightarrow\hspace{-1.5mm}^{}_B\rangle$ or a state orthogonal to it in the Hilbert space. 
So the result (\ref{ExpSigBz}) is equivalent to the one taken after a complete orthogonal measurement with the post-MS 
\begin{equation}
  \left\{ |\uparrow^{}_A \rightarrow\hspace{-1.5mm}^{}_B\rangle, |\uparrow^{}_A \leftarrow\hspace{-1.5mm}^{}_B\rangle, |\downarrow^{}_A \; \uparrow^{}_B\rangle,
  |\downarrow^{}_A \; \downarrow^{}_B\rangle \right\}. \label{PMSQNDSV}
\end{equation}
One can see that some nonlocal information has been put in by the QNDSV:
when the outcome of particle A is up ($|\uparrow^{}_A\rangle$), particle B {\it must} be in an eigenstate of $\sigma_B^x$, and
when the outcome of particle A is down ($|\downarrow^{}_A\rangle$), particle B can be in an eigenstate of $\sigma_B^z$ (or 
$\vec{\sigma}_B$ in any direction). It is this outcome-operator dependence producing a discrimination between $|\uparrow^{}_A\rangle$ 
and $|\downarrow^{}_A\rangle$ when Bob is taking $\langle\hat{s}_B^x \rangle$ locally. 
Similar post-MS have been applied to the discussions on the ``quantum nonlocality without entanglement" \cite{BDV99, GR02}. 

An R-nonlocal quantum state-verification measurement without nondemolition, either without fixing any of the post-MS, or 
with all of the orthgonal post-MS fixed, may still respect causality \cite{GR02}.
The lesson here is that if the number of the projection operators in describing a measurement process in a composite system
is less than the dimension of the Hilbert space of the total system, 
and at least one of the post-MS is fixed but not all, 
then we may be able to construct a protocol which violates causality using this description of measurement.

\section{Nonlocal variables of two spins}

Indeed, similar situations arise in the projective measurements of R-nonlocal variables with degenerate eigenstates.
Consider the same two-spin system held by Alice and Bob and write the total spin of the system as $\vec{S}=\vec{s}^{}_A + \vec{s}^{}_B$,
which is an R-nonlocal variable since particles A and B are separated at a distance.
For the operator $\hat{S^2} = \hat{\vec{S}}\cdot\hat{\vec{S}}$, the eigenstates corresponding to the eigenvalues $\hbar^2 S(S+1)$ are
\begin{eqnarray}
  S=0:  \hspace{.5cm} && {1\over \sqrt{2}} \left( { \over } | \uparrow^{}_A \downarrow^{}_B\rangle -  
      | \downarrow^{}_A \uparrow^{}_B\rangle { \over }\right),\nonumber\\
  S=1:  \hspace{.5cm} && {1\over \sqrt{2}} \left( { \over } | \uparrow^{}_A \downarrow^{}_B\rangle + 
      | \downarrow^{}_A \uparrow^{}_B\rangle { \over }\right), \;\;\;
  | \uparrow^{}_A \uparrow^{}_B\rangle, \;\;\;  | \downarrow^{}_A \downarrow^{}_B\rangle, \nonumber\\ && \;\;\;
  {\rm or \; their \; linear \; combinations,}
\label{eigenS2}
\end{eqnarray}
and for the operator $\hat{S}^z$, the eigenstates corresponding to the eigenvalues $\hbar m^{}_S$ are
\begin{eqnarray}
  m^{}_S=1:  \hspace{.5cm} && | \uparrow^{}_A \uparrow^{}_B\rangle, \nonumber\\
  m^{}_S=-1:  \hspace{.5cm} && | \downarrow^{}_A \downarrow^{}_B\rangle, \nonumber\\
  m^{}_S=0:  \hspace{.5cm} &&  | \uparrow^{}_A \downarrow^{}_B\rangle, \;\;\; | \downarrow^{}_A \uparrow^{}_B\rangle,
  \;\;\; {\rm or \; their \; linear \; combinations.}
\label{eigenSz}
\end{eqnarray}
Here both the variables $S^2$ and $S^z$ are Q-local because in the Hilbert space of the two-spin system  
the states living in the subspace spanned by the degenerate eigenstates of each variable (associated with $S=1$ or $m^{}_S=0$) 
are not affected by the measurement of that variable.

The eigenstates (\ref{eigenS2}) can be viewed as the post-MS for a QNDSV of the singlet state
$\left(|\uparrow^{}_A \downarrow^{}_B \rangle - |\downarrow^{}_A \uparrow^{}_B\rangle\right)/\sqrt{2}$. 
Thus it is not surprising that  
the ideal measurement of $\hat{S^2}$ can violate causality if particles A and B are separated at a distance. 
Below is an example given explicitly in \cite{BP99}. 
Suppose the system is prepared in the state $|\uparrow^{}_A \uparrow^{}_B\rangle$. 
If Alice chooses to do nothing on particle A before the measurement of $\hat{S^2}$ at $t=t_1$, then 
from (\ref{eigenS2}) the measurement of $\hat{S^2}$ will collapse the system to the post-MS
\begin{equation}
  |\uparrow^{}_A \uparrow^{}_B\rangle \hspace{.5cm} {\rm with\; probability \; 1},
\end{equation}
yielding $\langle \hat{s}_B^z\rangle = +\hbar/2$ right after $t_1$. On the other hand,
if Alice chooses to flip the spin of particle A right before $t_1$ so that the pre-MS becomes $|\downarrow^{}_A \uparrow^{}_B\rangle$, 
then the measurement will collapse the system to one of the post-MS,
\begin{eqnarray}
  && \left(| \uparrow^{}_A \downarrow^{}_B\rangle - | \downarrow^{}_A \uparrow^{}_B\rangle \right)/\sqrt{2} 
      \hspace{.5cm} {\rm with\; probability \; 1/2},\nonumber\\
  && \left( | \uparrow^{}_A \downarrow^{}_B\rangle + | \downarrow^{}_A \uparrow^{}_B\rangle \right)/\sqrt{2} 
      \hspace{.5cm} {\rm with\; probability \; 1/2},
\end{eqnarray}
which gives $\langle \hat{s}_B^z\rangle = 0$ right after $t_1$. Thus, by looking at $\langle \hat{s}_B^z\rangle$ right after the measurement,
Bob will be able to know whether Alice has flipped $\vec{s}_A$ right before $t_1$, no matter how far 
Alice and Bob are apart. This allows superluminal signaling and violates causality.

One resolution is to accept that the projective measurement of the R-nonlocal variable $S^2$ of two spatially separated spins
may simply be forbidden by the causality principle. 
To measure such kind of the R-nonlocal variables one may have to introduce an ancillary entangled pair of spatially localized qubits 
as probes to interact with particles $A$ and $B$ locally in space and time. 
After the local interactions were done, the information originally in particles A and B was carried by the probes 
which were then brought together and measured at the same spacetime point to retrieve the nonlocal information of the two-spin system 
\cite{AAV86, PV94}. It takes time to bring these information carriers together and so the whole process is causal.

While this resolution works well, an ambiguity in the composite systems leads us to an alternative resolution: Even if particles A and B are 
located at the same space point, once they are {\it distinguishable} particles with spins being the only degrees of freedom quantized, 
and once the pre-MS of them is not any eigenstate of a variable with degeneracy, the post-MS of the measurement of that variable 
can be ambiguous due to the degeneracy and the expectation values of certain operators taken after the measurement of that variable 
can be inconsistent. For example, if the pre-MS is $|\rightarrow\hspace{-1.5mm}^{}_A \uparrow^{}_B\rangle$, 
then after a projective measurement of $S^2$, the system may be collapsed to the following set of the post-MS, 
\begin{eqnarray}
  && \left(| \uparrow^{}_A \downarrow^{}_B\rangle \mp | \downarrow^{}_A \uparrow^{}_B\rangle\right)/\sqrt{2}  
      \hspace{.5cm} {\rm each\; with\; probability \; 1/4}, \nonumber\\
  && | \uparrow^{}_A \uparrow^{}_B\rangle  \hspace{.5cm} {\rm with\; probability \; 1/2}, \nonumber\\
  && | \downarrow^{}_A \downarrow^{}_B\rangle  \hspace{.5cm} {\rm with\; probability \; 0}, 
  \label{S2PMS1} 
\end{eqnarray}
or another possible set of the post-MS,
\begin{eqnarray}
  && \left(| \uparrow^{}_A \downarrow^{}_B\rangle \mp | \downarrow^{}_A \uparrow^{}_B\rangle\right)/\sqrt{2},  
      \nonumber\\
  && \left(| \uparrow^{}_A \uparrow^{}_B\rangle \mp | \downarrow^{}_A \downarrow^{}_B\rangle\right)/\sqrt{2}  
      \hspace{.5cm} {\rm each\; with\; probability \; 1/4}, \label{S2PMS2} 
\end{eqnarray}
since $\left( | \uparrow^{}_A \uparrow^{}_B\rangle \pm | \downarrow^{}_A \downarrow^{}_B\rangle \right)/\sqrt{2}$ are 
also eigenstates of $\hat{S^2}$ with $S=1$, as good as $| \uparrow^{}_A \uparrow^{}_B\rangle$ and $| \downarrow^{}_A \downarrow^{}_B\rangle$.
Both sets of the post-MS give $\langle \hat{S^2}\rangle = 3\hbar^2/2$ before and after the measurement as expected, 
but $\langle \hat{s}^z_B \rangle = \hbar/4$ for the post-MS in (\ref{S2PMS1}), which is inconsistent with the result 
$\langle \hat{s}^z_B \rangle = 0$ for (\ref{S2PMS2}) (both are inconsistent with $\langle \hat{s}^z_B \rangle = \hbar/2$ for the pre-MS, anyway). 
A similar ambiguity will still occur if the variable $\hat{S^2}$ is replaced by $\hat{S}^z$ in the above example
\footnote{For the same pre-MS $|\rightarrow\hspace{-1.5mm}^{}_A \uparrow^{}_B\rangle$, 
if the post-MS with $m_S=0$ are fixed to be $| \uparrow^{}_A \downarrow^{}_B\rangle$ and 
$| \downarrow^{}_A \uparrow^{}_B\rangle$, then $\langle \hat{s}_B^z \rangle = \hbar/2$ after the measurement of $S^z$. 
If they are $(| \uparrow^{}_A \downarrow^{}_B\rangle \pm | \downarrow^{}_A \uparrow^{}_B\rangle)/\sqrt{2}$ instead, then  
$\langle \hat{s}_B^z \rangle = \hbar/4$ after the measurement.
Similar ambiguity also happens to $\langle\hat{S^2}\rangle$ or $\langle\hat{S}^z\rangle$  after a naive ideal measurement of
$\hat{s}_A^z$ or $\hat{s}_B^z$.
}.
Such an ambiguity is relevant to the distinguishability of the particles and irrelevant to the R-nonlocality. 
To eliminate this ambiguity, one has to specify every element in the set of the post-MS 
even some of them have degenerate outcomes. 

For two spin-particles located at the same space point, it seems that no purely theoretical principle
but experimental results can tell which set of the post-MS would be the right choice.
Fortunately if the two particles can be separated in the position space, causality may serve as a guiding principle for theorists to select.
Indeed, one may insist both $\hat{S^2}$ and $\hat{S}^z$ are able to be described in von Neumann's prescription,  
and the post-MS of $\hat{S^2}$ must be one of the following states:
\begin{eqnarray}
  S=0:  \hspace{.5cm} && {1\over \sqrt{2}} \left( { \over } | \uparrow^{}_A \downarrow^{}_B\rangle -  
      | \downarrow^{}_A \uparrow^{}_B\rangle { \over }\right),\nonumber\\
  S=1:  \hspace{.5cm} && {1\over \sqrt{2}} \left( { \over } | \uparrow^{}_A \downarrow^{}_B\rangle +  
      | \downarrow^{}_A \uparrow^{}_B\rangle { \over } \right), \nonumber\\ &&
  {1\over \sqrt{2}} \left( { \over }| \uparrow^{}_A \uparrow^{}_B\rangle + | \downarrow^{}_A \downarrow^{}_B\rangle{ \over } \right), \;\;\;
  {1\over \sqrt{2}} \left( { \over }| \uparrow^{}_A \uparrow^{}_B\rangle - | \downarrow^{}_A \downarrow^{}_B\rangle{ \over } \right), 
\label{PMSS2}
\end{eqnarray}
with no linear combination of the above three states with $S=1$ being allowed, while 
the post-MS of $\hat{S}^z$ must be one of the following states:
\begin{eqnarray}
  m^{}_S=1:  \hspace{.5cm} && | \uparrow^{}_A \uparrow^{}_B\rangle, \nonumber\\
  m^{}_S=-1:  \hspace{.5cm} && | \downarrow^{}_A \downarrow^{}_B\rangle, \nonumber\\
  m^{}_S=0:  \hspace{.5cm} &&  | \uparrow^{}_A \downarrow^{}_B\rangle, \;\;\; | \downarrow^{}_A \uparrow^{}_B\rangle,
\label{PMSSz}
\end{eqnarray}
with all linear combinations of $| \uparrow^{}_A \downarrow^{}_B\rangle$ and $| \downarrow^{}_A \uparrow^{}_B\rangle$ being forbidden 
\cite{AAV86,BGNP01}. Then $\langle \hat{s}_B^z \rangle$ will be independent of Alice's local operation before the measurement.

In fact, (\ref{PMSS2}) and (\ref{PMSSz}) are the only two types of the post-MS set respecting causality in a two-spin system, as shown in \cite{PV94}.
Either (\ref{PMSS2}) or (\ref{PMSSz}) forms a complete set of the projection operators called the complete semicausal measurement 
superoperators in Ref. \cite{BGNP01}, where the criterion for such operators  
in a bipartite QM system to respect causality in the above setup has been given. For example,
when the state of particle A is traced out, all the four projection operators 
constructed by the four post-MS in (\ref{PMSS2}) yield the same reduced projection operators for particle B as ${1\over 2}{\bf 1}_B$. 
This satisfies the sufficient and necessary condition for the corresponding measurement being (semi-)causal.

Restricting the post-MS to be (\ref{PMSS2}) or (\ref{PMSSz}) eliminates the ambiguity of the post-MS with degenerate eigenstates
as well as the possibility of superluminal signaling in the R-nonlocal measurements.
In such a resolution quantum states in the whole Hilbert space will be affected by the R-nonlocal measurement.
In other words, the ideal measurements of R-nonlocal variables $S^2$ and $S^z$ have to be Q-nonlocal. 

Note that, in contrast to the conventional R-local measurement, if Bob has the knowledge about the initial state of the two-spin system, 
he will be able to know which kind of the R-nonlocal measurement has been performed by the external agent: for the initial state 
$| \uparrow^{}_A \uparrow^{}_B\rangle$, if the operator associated with the R-nonlocal measurement at $t_1$ was $\hat{S^2}$, then Bob would find 
$\langle \hat{s}_B^z \rangle=0$ at $t_1+\epsilon$; if it was $\hat{S}^z$, or there was no measurement has ever performed, then 
$\langle \hat{s}_B^z \rangle= +\hbar/2$. To respect causality, we have to further assume that the external agent has no information to send
to Bob (or Alice) in choosing the R-nonlocal measurement.


\section{Ideal measurement on two oscillators}

To get closer to the R-nonlocal projective measurement on bosonic fields, 
let us consider a pair of the point-like objects whose internal degrees of freedom $Q^{}_A$ and $Q^{}_B$ act as two 
quantum mechanical harmonic oscillators (HO) in 1D with the same natural frequency $\Omega$ and mass $m$. 
Again we assume these two dstinguishable objects are held by Alice and Bob separated in space. 
Analogous to the example in the previous section, we are looking at the post-MS of the measurements of the scaled
``center of mass" operator $Q_+$ and ``relative position" operator $Q_-$ defined by
\begin{equation} 
\hat{Q}_\pm \equiv {\hat{Q}^{}_A \pm\hat{Q}^{}_B\over\sqrt{2}},
\end{equation} 
as well as $\hat{P}_\pm \equiv (\hat{P}^{}_A \pm\hat{P}^{}_B)/\sqrt{2}$ where $P_j$ is the conjugate momentum of $Q_j$.

Suppose the two-HO system is initiated in a product of the coherent states 
\begin{eqnarray}
\psi_{\rm in} (Q^{}_A, Q^{}_B) &=& e^{i p^{}_A Q^{}_A/\hbar}\psi_0(Q^{}_A) e^{i p^{}_B Q^{}_B/\hbar} \psi_0(Q^{}_B)\nonumber\\ 
  &=& \left[ e^{i p^{}_+ Q^{}_+/\hbar} \psi_0(Q_+)\right]\left[ e^{i p^{}_- Q^{}_-/\hbar}\psi_0(Q_-)\right]
\end{eqnarray} 
before $t_1-\varepsilon$. Here $p^{}_\pm \equiv (p^{}_A \pm p^{}_B)/\sqrt{2}$ are real-number parameters and 
$\psi_n(z) \equiv (\kappa/\sqrt{\pi} 2^n n!)^{1/2} H_n(\kappa z)e^{-\kappa^2 z^2/2}$ are energy eigenstates of HO with 
$\kappa \equiv \sqrt{m\Omega/\hbar}$, $n=0,1,2,3,\cdots$, and $H_n(x)$ the Hermite polynomials.
At $t_1-\varepsilon$, Alice gives a ``kick" to $Q^{}_A$ 
by applying a unitary operator $\exp \{i \hat{Q}^{}_A \lambda /\hbar \}$ to the initial state, 
i.e., shift $P^{}_A$ by amount $\lambda$ freely chosen by Alice, 
then the pre-MS right before the measurement reads
\begin{equation}
  \psi_{M_-} = e^{i \lambda Q^{}_A /\hbar} \psi_{\rm in}(Q^{}_A, Q^{}_B) = 
    \left[ e^{i\lambda_+ Q_+/\sqrt{2}\hbar}\psi^{}_0(Q_+)\right] \left[ e^{i\lambda_- Q_-/\sqrt{2}\hbar}\psi^{}_0(Q_-)  \right]
\label{psiinHO}
\end{equation}
with $\lambda_\pm \equiv p^{}_A \pm p^{}_B + \lambda$.

Consider an R-nonlocal measurements of the number operator $\hat{N}_+$ for $Q_+$, where
$\hat{N}^{}_j \equiv \hat{a}_j^\dagger \hat{a}^{}_j$ with the lowering operators
\begin{equation}
   \hat{a}_j = \sqrt{m\Omega\over 2\hbar} \hat{Q}_j + i \sqrt{1\over 2\hbar m\Omega} \hat{P}_j, \hspace{.5cm} j=A,B,+,-,
\end{equation}
commutes with the Hamiltonian of the two HO.                                                                                             
Naively, since (\ref{psiinHO}) is a separable state in terms of $Q_+$ and $Q_-$, one may follow the same rule in the R-local measurements 
\cite{Lin12, LSCH12} that only the $Q_+$ part of the pre-MS (\ref{psiinHO}) is projected so that the post-MS has the form
\begin{equation}
 \psi^{}_n(Q_+)\psi^{}_0(Q_-) e^{i\lambda_- Q_-/\sqrt{2}\hbar} 
 \label{naivePMS}
\end{equation} 
for some non-negative integer $n$.
(Below we call the Q-local measurements corresponding to this kind of the post-MS
as the {\it naive} ideal measurements of R-nonlocal variables.) Then (\ref{naivePMS}) gives 
$\langle \hat{P}^{}_B\rangle = -(p^{}_A-p^{}_B+\lambda)/2$, 
which depends on $\lambda$ chosen by the free will of Alice and so violates causality. 

Learning from the previous sections,  
to eliminate the dependence of $\lambda$ and get rid of the superluminal signaling,
one may further restrict the post-MS to be one of the complete orthonormal basis 
$\psi^{}_{\alpha\beta} = \psi^{}_\alpha(Q_+)\varphi^{}_\beta(Q_-)$ with some quantum numbers $\alpha$ and $\beta$.
Then the pre-MS can be expanded as
\begin{equation}
  \psi_{M_-}(Q^{}_A,Q^{}_B) = \sum_{\alpha,\beta} \, c^{}_{\alpha\beta}(\lambda)\; \psi_{\alpha\beta}(Q^{}_A,Q^{}_B)
  \label{Psiinalbe}
\end{equation}
with all the $\lambda$-dependence contained in the coefficient $c^{}_{\alpha\beta}$, and the expectation values of Bob's
local variables $f(\hat{Q}^{}_B, \hat{P}^{}_B)$ after measurement reads
\begin{equation} 
  \langle f(\hat{Q}^{}_B, \hat{P}^{}_B) \rangle = \sum_{\alpha, \beta} \left| c^{}_{\alpha\beta}(\lambda) \right|^2
  \langle f(\hat{Q}^{}_B, \hat{P}^{}_B) \rangle_{\alpha\beta}
\end{equation}
where
\begin{equation} 
  \langle f(\hat{Q}^{}_B, \hat{P}^{}_B) \rangle_{\alpha\beta}\equiv 
  \int dQ^{}_A dQ^{}_B \,\psi^*_{\alpha\beta}(Q^{}_{A,B}) f(\hat{Q}^{}_B, \hat{P}^{}_B)\psi^{}_{\alpha\beta} (Q^{}_{A,B}).
\label{HOcausalCond}
\end{equation}
To make $\langle f(\hat{Q}^{}_B, \hat{P}^{}_B) \rangle$ independent of $\lambda$, one possibility is that 
$\langle f(\hat{Q}^{}_B, \hat{P}^{}_B) \rangle_{\alpha\beta}$ are evenly distributed in positive and negative values,
as occured in the two-spin case with the post-MS (\ref{PMSSz}) and the operator $f\sim \hat{s}^z_B$. 
A simpler situation is that for every operator $f(\hat{Q}^{}_B, \hat{P}^{}_B)$ Bob has a constant
finite value of $\langle f(\hat{Q}^{}_B, \hat{P}^{}_B) \rangle_{\alpha\beta}$ independent of $\alpha$ and $\beta$ 
whenever $c^{}_{\alpha\beta}(\lambda)\not=0$, as in the two-spin case with the post-MS (\ref{PMSS2}) and the operator $f\sim \hat{s}^z_B$.
The $\lambda$-dependence on Bob's side is erased
since $\sum_{\alpha,\beta} |c^{}_{\alpha\beta}(\lambda)|^2=1$ after the constant $\langle f\rangle_{\alpha\beta}$ is factored out of the sum.

Unfortunately in the ideal measurement of $\hat{N}_+$ on the two-HO system both the above two situations are not easy to achieve. 
For example, if one chooses the post-MS as the basis $\psi^{}_{nn'} = \psi^{}_n(Q_+)\psi^{}_{n'}(Q_-)$, while the expectation values 
$\langle \hat{Q}^{}_B \rangle^{}_{nn'} = \langle \hat{P}^{}_B \rangle^{}_{nn'} =0$, one has 
$\langle \hat{Q}^{2}_B \rangle^{}_{nn'}$ and $\langle \hat{P}^{2}_B \rangle^{}_{nn'}$ proportional to $n+n'+1$,
so that $\langle \hat{Q}^{2}_B \rangle$ and $\langle \hat{P}^{2}_B \rangle$ depend on $\lambda$ explicitly.

An alternative strategy is to make those $\langle f(\hat{Q}^{}_B, \hat{P}^{}_B) \rangle_{\alpha\beta}$ which depends on $\alpha$ or $\beta$
diverge to wash out the $\lambda$-dependence.
This suggests that if $\psi^{}_\alpha(Q_+)$ is an eigenstate of the R-nonlocal variable $g(\hat{Q}_+, \hat{P}_+)$, then 
$\varphi^{}_\beta(Q_-)$ can be something like an eigenstate of the variable conjugate to $g$.
Indeed, for our R-nonlocal measurement of $\hat{N}_+$, one may choose the post-MS to be either 
\begin{equation}
    \psi_n(Q_+) \chi_{\theta_s}^{(0)}(Q_-) \hspace{1cm} {\rm or} \hspace{1cm} \psi_n(Q_+) \chi_{\theta_s}^{(1)}(Q_-) 
\end{equation}
for some $n=0,1,2,3,\ldots,S$ and $\theta_s \in [0, 2\pi)$. Here 
\begin{equation}
  \chi^{(b)}_{\theta_s} (Q_-) \equiv \lim_{S\to\infty}{1\over \sqrt{S+1}} \sum_{n=0}^S e^{i (2n+b)\theta_s} \psi_{2n+b}(Q_-),
  \hspace{.5cm}\theta_s = {2\pi s\over S+1},
\label{chiphase}
\end{equation}
with $s=0,1,2,3,\ldots,S$ are the parity-even ($\chi^{(0)}_{\theta_s}$) and parity-odd ($\chi^{(1)}_{\theta_s}$)
parts of the phase state $\chi$ $(= \chi^{(0)}+ \chi^{(1)})$ \cite{PB97}.
Then the excitation number of $Q_-$ becomes totally uncertain  
and the coefficients $c^{}_{\alpha\beta}(\lambda)$ in the expansion (\ref{Psiinalbe}) read
\begin{equation}
  c^{}_{n\{b\theta_s \}} = e^{-(\Lambda_+^2 +\Lambda_-^2)/2} { (i\Lambda_+)^{n} \over\sqrt{n!}} \lim_{S\to\infty} \sum_{j=0}^{S} 
  {\left(i \Lambda_- e^{-i\theta_s}\right)^{2j+b} \over \sqrt{(2j+b)! (S+1)}}
\end{equation}
with $\Lambda_\pm \equiv \lambda_\pm/2\hbar\kappa$, and right after the measurement, 
\begin{equation}
  \langle \hat{Q}_B \rangle = \lim_{S\to\infty}\sum_{n=0}^{2S+1}\sum_{b=0}^1 \sum_{s=0}^S \left| c^{}_{n\{b\theta_s \}}\right|^2 
  \langle \hat{Q}_B \rangle_{n \{b\theta_s\}} =0
\end{equation}
since $\langle \hat{Q}_B \rangle_{n \{b\theta_s\}} 
= 0$ for all $n$, $\theta_s$, and $b$. Similarly $\langle \hat{P}_B \rangle =0$, while
both $\langle \hat{Q}_B^2 \rangle$ and $\langle \hat{P}_B^2 \rangle \sim \lim_{S\to \infty} S$ diverge.
Thus the $\lambda$-dependence cannot be observed by Bob.

Again, all the local information initially in $Q^{}_A$ and $Q^{}_B$ will be eliminated by the above R-nonlocal measurement, 
and similar to the two-spin cases, now Bob can know whether the R-nonlocal measurement of $Q_+$ has been performed or not
by examining $\langle \hat{Q}^{2}_B\rangle$ being infinity (measured) or finite (not yet) in the above example.

The above R-nonlocal measurement of $\hat{N}_+$ would introduce a huge impact making the energy of oscillator B, 
$\langle E_B \rangle = \langle (\hat{P}_B^2 / 2m) + (m\omega^2 \hat{Q}_B^2/2)\rangle $, to blow up after the measurement. 
This suggests that such a measurement costs infinitly large energy for the external agent.
The infinite expectation values in (\ref{HOcausalCond}) will become finite
if one assigns a finite value to the upper-limit $S$ in (\ref{chiphase}),  but then
causality will be formally violated by the explicit $\lambda$-dependence. 
We may have to accept that in realistic systems consisting of localized HO, ideal measurements of R-nonlocal variables  
would always violate causality formally up to the level determined by the cutoffs of the theory. 
When formulating an effective theory for a system of this kind, one has to make sure that no causality-violating signal can be resolved 
or observed by any element or apparatus in the range of validity of the theory. 

\section{Ideal measurement of particle number of a scalar field}
\label{IdealField}

Now we are ready to consider the ideal measurement of R-nonlocal variables in RQFT.
Suppose we are measuring the particle number $\hat{N}_{\bf p}$ of some mode ${\bf p}$ of a real scalar field
initially in a state generated from the Minkowski vacuum. From (\ref{Minkvac}) and (\ref{1particle})
one can see that usually $\phi_{\bf k}$ and $\phi_{-\bf k}$ are coupled together in this kind of states. 
So the post-MS after measuring $\hat{N}_{\bf p}$ would be in the form
\begin{equation}
  \psi^{}_{mn} (\phi_{\bf p},\phi_{-\bf p}) \varphi[\phi_{{\bf k}\not=\pm{\bf p}}] 
\label{PMSQFT}
\end{equation}
where
\begin{equation}
   \psi^{}_{mn}(\phi_{\bf p}, \phi_{-\bf p})  = \sqrt{ \omega^{}_{\bf p}\epsilon \over m! n! \hbar\pi}
   \left(\hat{b}^\dagger_{\bf p}\right)^m  
     \left(\hat{b}^\dagger_{-\bf p}\right)^n\exp -{\epsilon\over \hbar} \omega_{\bf p} \phi^{}_{\bf p} \phi^{}_{-{\bf p}}
\end{equation}
is $m$- and $n$-particle state of the field modes ${\bf p}$ and $-{\bf p}$, respectively. Here 
$\epsilon \equiv 1/[(2\pi)^3 \delta^3(0)] \sim d^3 k/(2\pi)^3$ is the volume element of the wave-vector space and 
the creation operator of the field mode is defined by \cite{LCH09}
\begin{equation}
  \hat{b}_{\pm{\bf p}}^\dagger = \sqrt{\omega_{\bf p}\epsilon\over 2\hbar} \, \phi^{}_{\mp\bf p} - \sqrt{\hbar\over 2\omega_{\bf p}\epsilon}
    \,{\partial\over \partial \phi^{}_{\pm{\bf p}}}.
\end{equation}
From the experience in the two-HO cases, one is tempted to further restrict $\varphi[\phi_{{\bf k}\not=\pm{\bf p}}]$ in (\ref{PMSQFT})
to be $\prod_{{\bf k}\not=\pm{\bf p}}\varphi(\phi_{\pm\bf k})$, 
$\varphi(\phi_{\pm\bf k}) \sim \chi^{(a)}_{\theta_{\pm{\bf k}}}(\phi_{\pm{\bf k}})$ to respect causality. 
This may cost infinite energy and make the particle numbers of all modes but $\pm{\bf p}$ totally uncertain after the measurement.
Fortunately, our argument in previous sections about the ambiguity in the QM composite systems is for distinguishable particles or atoms, 
so those considerations may not apply to RQFT. 
More interestingly, when we apply the naive post-MS similar to (\ref{naivePMS}) to RQFT, 
the situation turns out to be much better than those in the quantum mechanical two-HO and two-spin cases. 

Consider a setup similar to Sorkin's example except that the QNDSV in Sec. \ref{SorkinEx} will be replaced by the naive ideal measurement
described below.
Suppose initially the field is in the Minkowski vacuum (\ref{Minkvac}), and after Alice performed the R-local operation 
${\bf U}_{\bf x}$ in (\ref{Ux}), $\langle \pi_{\bf x} \rangle$ is shifted by amount $\lambda \delta^3(0)$ and 
the wave functional of the pre-MS reads
\begin{eqnarray}
  & &\Psi_{M_-} = {\bf U}_{\bf x} \Psi_0 \nonumber\\ &=&
  {\cal N} e^{-i E_0 t_1/\hbar} \exp \left[
  -{1\over 2\hbar}\int d^3 z d^3 z' \phi_{\bf z} g({\bf z}-{\bf z}') \phi_{{\bf z}'} -{i\over\hbar}\lambda\phi_{\bf x}\right]\nonumber\\
  &=& {\cal N} e^{-i E_0 t_1/\hbar} \exp 
  {-1\over 2\hbar}\int d^3 k \left[ \omega_{\bf k} \phi_{\bf k}\phi_{-{\bf k}} + i \lambda \left( 
  e^{i{\bf k}\cdot{\bf x}} \phi_{\bf k} +  e^{-i{\bf k}\cdot{\bf x}} \phi_{-{\bf k}}\right)\right].
\label{preMS}
\end{eqnarray} 
Suppose a naive ideal measurement of the particle number of a specific mode-pair $\pm{\bf p}$ is performed at $t_1$.
Since $\Psi_{M_-}$ is separable in terms of $\phi_{\pm {\bf k}}$ for each mode-pair $\pm {\bf k}$, 
the field state will collapse to a post-MS similar to (\ref{naivePMS}),
\begin{eqnarray}
  & &\Psi_{mn}[\phi] \nonumber\\ &=& \psi^{}_{mn}(\phi_{\bf p}, \phi_{-\bf p}) \int d\phi'_{\bf p}\, d\phi'_{-\bf p}  
    \psi^*_{mn}(\phi'_{\bf p}, \phi'_{-\bf p})  \Psi^{}_{M_-}[\phi'_{\bf p}, \phi'_{-\bf p}, \phi^{}_{{\bf k}\not=\pm{\bf p}}]\nonumber\\
  &=& {e^{i(n-m){\bf p}\cdot {\bf x}} \over \sqrt{m! n!}}\left[i \lambda 
  \sqrt{\omega^{}_{\bf p}\epsilon\over 2\hbar}\right]^{m+n} 
    e^{- \lambda^2 \omega^{}_{\bf p}\epsilon /2\hbar} \psi^{}_{mn}(\phi_{\pm\bf p}) 
    {\prod_{{\bf k}\not=\pm{\bf p}}} \varphi(\phi^{}_{\pm\bf k}), 
    \label{naivePMSQFT}
\end{eqnarray}
up to a normalization factor, for some $m$, $n \in {\bf Z}$ if the outcome of the particle number of the mode $\pm {\bf p}$ 
is $m$ particles with ${\bf p}$ and $n$ particles with $-{\bf p}$. Here the field modes $\phi_{{\bf k} \not= \pm {\bf p}}$ 
are not affected by the measurement and thus, the same as the pre-MS, one has
$\varphi(\phi_{\pm\bf k}) 
= \left(\omega^{}_{\bf k}\epsilon/\hbar\pi\right)^{1/4} e^{ -\epsilon/2\hbar} [ 
    \omega_{\bf k} \phi^{}_{\bf k} \phi^{}_{-{\bf k}} - i\lambda \left( e^{i {\bf k}\cdot{\bf x}} \phi^{}_{\bf k}+ 
    e^{-i {\bf k}\cdot{\bf x}} \phi^{}_{-\bf k}\right)  ]$
in (\ref{naivePMSQFT}).
Note that the above $\Psi_{mn}[\phi]$ is not normalized, and the probability of finding the post-MS being proportional to $\Psi_{mn}[\phi]$
is exactly $P_{mn} \equiv\int {\cal D}\phi \Psi_{mn}^*[\phi] \Psi_{mn}[\phi]$
(with ${\cal D}\phi =  \prod_{\bf k} d\phi_{\bf k}$ or $\prod_{\bf z} d\phi_{\bf z}$, depending on the representation;
It is straightforward to verify $\sum_{m,n=0}^\infty P_{mn}=1$). Analogous to the situation in (\ref{result1}),
$P_{mn}$ cancels the normalization factor of the post-MS in the calculation so that the expectation value of $\phi_{\bf y}$ found by Bob is
\begin{eqnarray}
  \langle \phi^{}_{\bf y} \rangle &=& \sum_{m,n=0}^\infty \int {\cal D}\phi  \Psi^*_{mn}[\phi] \int {d^3 k\over (2\pi)^3} 
    e^{i {\bf k}\cdot {\bf y}} \phi_{\bf k}  \Psi^{}_{mn}[\phi] \nonumber\\
  &=& \sum_{m,n=0}^\infty \int d\phi_{\bf p} d\phi_{-{\bf p}}\psi^*_{mn}(\phi_{\pm\bf p}) \epsilon \left[ 
    e^{i\bf p\cdot \bf y}\phi_{\bf p}+ e^{-i\bf p\cdot \bf y}\phi_{-\bf p}\right] \psi^{}_{mn}(\phi_{\pm\bf p})\nonumber\\ &+&  
    {1\over 2}\int_{{\bf k}\not=\pm {\bf p}} {d^3 k\over (2\pi)^3} \int d\phi_{\bf k} d\phi_{-\bf k} 
    \varphi^*(\phi_{\pm\bf k}) \left[ e^{i {\bf k}\cdot {\bf y}}\phi_{\bf k} + e^{-i {\bf k}\cdot {\bf y}}\phi_{-\bf k}\right] 
    \varphi(\phi_{\pm\bf k}) \nonumber\\
  &=& 0.
\end{eqnarray}
Similar calculations give
\begin{eqnarray}
  \langle \pi^{}_{\bf y} \rangle &=& \lambda\left[ \delta^3({\bf x}-{\bf y}) -
      2 \epsilon \cos {\bf p}\cdot({\bf x}-{\bf y}) \right], \label{Naivepi}\\
  \langle \phi^{2}_{\bf y} \rangle &=& \int {d^3 k\over (2\pi)^3}{\hbar\over 2\omega_{\bf k}} + 
      \lambda^2 \epsilon^2,\\
  \langle \pi^{2}_{\bf y} \rangle &=& \langle \pi^{}_{\bf y} \rangle^2 + \int {d^3 k\over (2\pi)^3}{\hbar\over 2}\omega_{\bf k} + 
      \left( \lambda \epsilon \omega_{\bf p} \right)^2 ,
\end{eqnarray}
where $\pi^{}_{\bf y}$ is the conjugate momentum of $\phi^{}_{\bf y}$.
One can see that, while there are still $\lambda$-dependence in these results for ${\bf y}$ far apart from ${\bf x}$,
every $\lambda$-dependent term for ${\bf y}\not={\bf x}$ is suppressed by a factor $\epsilon = 1/[(2\pi)^3\delta^3(0)] = 1/[(2\pi)^3 V]$, 
where $V$ is the total spatial volume of the background spacetime that the scalar field is living in. 
Moreover, the $\lambda$-dependent terms in $\langle \phi^{2}_{\bf y} \rangle$ and $\langle \pi^{2}_{\bf y} \rangle$ are 
infinitesimal corrections to infinitely large values.
By letting $V$ go to infinity, or equivalently, letting the IR cutoff go to zero, the $\lambda$-dependent terms in the above results becomes 
negligible, and we reach the conclusion that the RQFT in Minkowski space with the naive ideal measurement included
is effective causal. 
In a field theory defined in a background space compact and small, however, causality violation by 
the above $\lambda$-dependent terms may be significant.

One obvious reason to have such tiny $\lambda$-dependent terms is that we only perform the measurement on one {\it single} mode-pair with
$\pm{\bf p}$, whose volume element in the ${\bf k}$-space is $\epsilon \sim d^3 k \sim 1/V$ only. 
We will still have the same suppression if the naive ideal measurement is performed on a single mode of orthogonal wave packets 
with continuous spectrum \cite{SR07}, though those wave packets are superpositions of plane waves. 
It seems that if our measurement is performed simultaneously on many modes around $\pm {\bf p}$ with a bandwidth $\Delta p$ 
(recall that $\Psi_{M_-}$ is a product state in terms of $\phi_{\pm{\bf k}}$), 
we could have larger finite $\lambda$-dependent corrections ($\sim (\Delta p)^3$). Nevertheless, 
any finite corrections to the infinitely large quantity $\langle \phi^{2}_{\bf y} \rangle $ or 
$\langle \pi^{2}_{\bf y} \rangle - \langle \pi^{}_{\bf y} \rangle^2$ are harmless, and the $\lambda$-dependent term in 
$\langle \pi^{}_{\bf y} \rangle$ with ${\bf y}\not= {\bf x}$ may be further suppressed when summing the oscillating function
$\cos {\bf k}\cdot({\bf x}-{\bf y})$ over ${\bf k}$ around $\pm{\bf p}$ (see (\ref{Naivepi})). 
The larger $|{\bf x}-{\bf y}|$, the stronger suppression is expected.

Now we see that RQFT is more tolerant to the naive ideal measurement of R-nonlocal variables than QM is. 
How about the QNDSV in RQFT?

\section{QNDSV of 1-particle state revisited}
\label{revisitQNDSV}

Let us revisit Sorkin's example. Again, the pre-MS right before $t_1$ is (\ref{preMS}).
At $t_1$, a QNDSV of the one-particle state (\ref{1particle}) of mode ${\bf p}$ is performed. 
Then right after $t_1$, the expectation value (\ref{result1}) observed by Bob has an explicit form
\begin{eqnarray}
  \langle \phi_{\bf y} \rangle 
  &=& -2 {\rm Re}\left\{  \langle {M_-}  | \hat{\phi}_{\bf y} | 1_{\bf p} \rangle \langle 1_{\bf p}| {M_-} \rangle \right\} \nonumber\\
  &=& -2 {\rm Re}\left\{  \int {\cal D}\phi \Psi^*_{M_-} \phi_{\bf y} \Psi_{1_{\bf p}} \int {\cal D}\phi' 
    \Psi^*_{1_{\bf p}} \Psi_{M_-} \right\}\nonumber\\
  &=& -\lambda \left(\exp  -{\lambda^2\over 4\hbar}  g^{-1}_{{\bf xx}} \right)^2
    \left( 2\omega_{\bf p} \epsilon \over \hbar \right)  \times \nonumber\\ & &{\rm Re} \, i\int d^3 z d^3 z' 
    e^{-i{\bf p}\cdot ({\bf z} - {\bf z}')} g^{-1}_{{\bf zy}} \left[ {\hbar\over 2} g^{-1}_{{\bf xz}'}
    -{\lambda^2\over 4}g^{-1}_{{\bf xy}} g^{-1}_{{\bf yz}'} \right] ,  \label{result1g} \\
  &=& \lambda  e^{-\lambda^2 g^{-1}_{{\bf xx}}/2\hbar} \left({2\epsilon \over \omega_{\bf p}}\right)
    \sin{\bf p}\cdot({\bf y}-{\bf x}) ,  
\label{result1rev}
\end{eqnarray}
where
\begin{equation}
  g^{-1}_{{\bf xy}}\equiv g^{-1}({\bf x}-{\bf y}) = \int{d^3 k\over (2\pi)^3}{1\over \omega_{\bf k}} 
    e^{-i {\bf k}\cdot({\bf x}-{\bf y})}
\end{equation}
denotes the inverse function of $g_{\bf xy}\equiv g({\bf x}-{\bf y})$ so that $g^{-1}_{\bf xy}g^{\bf{yz}}\equiv
\int d^3 y \left. g^{-1}({\bf x}-{\bf y})\right.$ $\left.g({\bf y} - {\bf z})\right.
 = \delta^3({\bf x}-{\bf z}) \equiv \delta_{\bf x}{}^{\bf z}$,
and $g^{-1}_{\bf xx} \equiv \lim^{}_{{\bf x'} \to {\bf x}} g^{-1}_{\bf x x'}$.
Obviously Bob's $\langle \phi_{\bf y} \rangle$ depends on Alice's parameter $\lambda$ of her local operation, 
no matter how far Alice and Bob are separated apart. 

If we generalize the one-particle state of a field mode in the QNDSV by an one-particle state of the wave packet (\ref{1particleWP}), 
we will have
\begin{eqnarray}
  \langle \phi_{\bf y} \rangle 
  &=& \lambda e^{-\lambda^2 g^{-1}_{{\bf xx}}/2\hbar} S({\bf x},{\bf y}),
\label{resultwavepk}
\end{eqnarray}
where
$S({\bf x},{\bf y})\equiv {\rm Im}\, {\cal F}^*(t_1, {\bf x}){\cal F}(t_1,{\bf y})$
with
\begin{equation}
  {\cal F}(t, {\bf z}) \equiv  \int {d^3 k\over (2\pi)^3} {\tilde{\varphi}({\bf k}) \over \sqrt{\omega^{}_{\bf k}}}  
  e^{-i(\omega^{}_{\bf k} t - {\bf k}\cdot {\bf z})} .
\end{equation}
Inserting $\tilde{\varphi}({\bf k}) = \sqrt{\epsilon} (2\pi)^3 \delta^3({\bf k}-{\bf p})$,
one recovers the result for the single mode ${\bf p}$ in (\ref{result1rev}).
Once we keep $g^{-1}_{\bf xx}$ finite by introducing cutoffs, 
we will reach Sorkin's result (Eq.(4) with $\beta=0$ in \cite{So93}) 
\begin{equation}
 \left. {d\langle \phi_{\bf y} \rangle\over d\lambda}\right|_{\lambda \to 0}  = S({\bf x},{\bf y})
\label{SorkinResult}
\end{equation}
which will not vanish if $S({\bf x},{\bf y})$ does not. This is why
the quantity $S({\bf x},{\bf y})$ plays the central role in \cite{BBBD12}.

Nevertheless, $g^{-1}_{\bf xx}$ in the overall factor $\exp \{-\lambda^2 g^{-1}_{{\bf xx}}/2\hbar \}$ in 
(\ref{result1rev}) and (\ref{resultwavepk}) formally diverges when integrated over the whole $k$-space 
(a consequence of the R-locality of the kick $U_{\bf x}$ in (\ref{Ux})),
so $\langle \phi_{\bf y} \rangle$ in (\ref{result1rev}) is strongly suppressed.
The $\lambda$-dependent terms in the expectation values of other R-local operators 
at Bob's position ${\bf y}$ also have the same suppression factor after this QNDSV, e.g.,
\begin{eqnarray}
  \langle \phi_{\bf y}^2 \rangle &=& {3\hbar\over 2}g^{-1}_{\bf yy} + {2\hbar \epsilon\over 
    \omega_{\bf p}} - \nonumber\\ & &{ \lambda^2 \epsilon \over \hbar 
    \omega_{\bf p} }e^{ -\lambda^2 g^{-1}_{{\bf xx}}/2\hbar} 
    \left[ 2\hbar g^{-1}_{\bf xy} \cos{\bf p}\cdot({\bf x}-{\bf y}) + {\hbar\over 2}g^{-1}_{\bf yy} - 
    {\lambda^2\over 4}\left(g^{-1}_{\bf xy}\right)^2  \right].
\label{result2}
\end{eqnarray}
Again, the $\lambda$-dependent term here is an infinitesimal correction to an infinitely large value.

When $\lambda^2 = \hbar/g^{-1}_{{\bf xx}}$, the combination $\lambda \exp  (-\lambda^2 g^{-1}_{\bf xx}/2\hbar)$ in (\ref{result1rev}) and (\ref{resultwavepk}) reach the maximum value $(\hbar/e^1 g^{-1}_{\bf xx})^{1/2}$. 
One may say causality is respected down to the order of
$(\hbar/e^1 g^{-1}_{\bf xx})^{1/2}$, which depends on the UV cutoff of the field in (3+1)D Minkowski space 
and/or the spatial resolution of the apparatus. For single modes, causality violation of $\langle \phi_{\bf y} \rangle$
in (\ref{result1rev}) is further suppressed by the factor $\epsilon$, which depends on the IR cutoff.
By letting the UV cutoff go to infinity and the IR cutoff go to zero, we conclude that the QNDSV of an one-particle state of a field mode 
in RQFT respects causality effectively.

\section{Summary}

We have studied the nonlocal projective measurements in the QM systems with two localized objects held by Alice and Bob 
separated in the position space, and in a RQFT system where Alice and Bob can operate the local degrees of freedom at their own positions.
The general setting is that Alice performs a local operation on one element located at her position right before the nonlocal measurement on the composite system, then Bob observes the expectation values of another element located at his position right after the nonlocal measurement.

In QM, naive extensions of the von Neumann's projective measurement 
can produce superluminal signaling in this setup and so violate causality both in the QNDSV and the ideal measurements of R-nonlocal variables.
We showed that in a QM two-spin system, the QNDSV violates causality even if both the pre-MS and the to-be-verified state 
of the two spins are separable states in terms of the R-local variables. Such causality violation is due to the ``nonlocality without entanglement",
which is possible when the number of projection operators in the process is less than the dimension of the Hilbert space.

The ideal measurements of R-nonlocal variables in the two-spin or two-HO QM system can be made causal by introducing 
further restrictions on the post-MS, so that the expectation values of all combinations of Bob's local dynamical variables for each post-MS 
are either zero or infinity. 
This makes the measurement Q-nonlocal and eliminates all the information originally in 
the pre-MS before measurement, so that Bob can know whether the R-nonlocal measurement was done or not from his observation.
Nevertheless, in realistic systems consisting of spatially localized HO, causality under the ideal measurements of R-nonlocal variables 
would only be respected up to some level determined by the cutoffs of the theory. 
In RQFT the situation appears similar. Causality violation also arises formally in a similar setup, and could be eliminated 
by introducing further restrictions on the post-MS. 
However, the causality-violating terms in Bob's results obtained right after a naive R-nonlocal but Q-local
ideal measurement on a single mode are suppressed by the IR cutoff of the field theory, and thus virtually undetectable. 
Causality violation by the QNDSV in RQFT is further suppressed by the UV cutoff.
It turns out that RQFT is more tolerant to the naive ideal measurement of R-nonlocal variables than QM is. 
A field theory with the quantum field defined in an infinitely large spacetime will be effectively
causal with the naive R-nonlocal projective measurement included.

Recall that in classical electrodynamics there exist acausal solutions for the motion of a point charge
in electromagnetic field \cite{Jackson}. However, once the length-scale is no smaller than the classical electron radius or 
the time-resolution is no shorter than $\tau_0 \equiv (2e^2/3mc^3)$ \cite{JH05}, formal causality violation will not be observed 
and so the theory is effectively causal  
\footnote{One should be cautious here. When looking at the {\it point charge} in classical electrodynamics, 
the UV cutoff cannot be too high to include the acausal effects, and there is no requirement on the IR cutoff. 
For the QNDSV of the {\it field mode} in Section \ref{revisitQNDSV}, however, the UV cutoff has to be high enough,
and for the ideal measurement in Section \ref{IdealField}, the IR cutoff has to be low enough to suppress causality violation.}.
For quantum fields, the same idea applies. 
In the cases studied in the present paper, infinitely many degrees of freedom and continuous spectrum 
of RQFT help to protect causality in effective theories, though the setup looks similar to the counterpart in QM
which suffers significant superluminal signals.  \\ \\

\noindent{\bf Acknowledgment}   
I thank Jason Doukas and Andrzej Dragan 
for helpful discussions. 
This work is supported by the National Science Council of Taiwan under grant NSC 102-2112-M-018-005-MY3, 
and in part by the National Center for Theoretical Sciences, Taiwan.\\ \\

\begin{appendix}

\section{Wave-functionals for vacuum and one-particle states of a scalar field}

The Fock states of a field at each moment are defined on the whole time-slice in some reference frame, and is entangled in terms of 
$\phi({\bf x}) \equiv \phi_{\bf x}$. 
For example, for a real scalar field $\phi$ with mass $m$ 
in (3+1)D Minkowski space, the wave-functional for the Minkowski vacuum can be written as \cite{Hat92}
\begin{eqnarray}
\Psi_{0} [ \phi; t ] &=& \langle\phi |e^{-i\hat{H}t/\hbar} |0\rangle = {\cal N} e^{-i E_0 t/\hbar}
  \exp -{1\over 2\hbar}\int {d^3 k\over (2\pi)^3}\omega_{\bf k} \phi_{\bf k}\phi_{\bf -k} \nonumber\\ &=& 
  {\cal N} e^{-i E_0 t/\hbar} \exp -{1\over 2\hbar}\int d^3 z d^3 z' \phi_{\bf z} g({\bf z}-{\bf z}') \phi_{{\bf z}'}  
\label{Minkvac}
\end{eqnarray}
where $\hat{H}$ is the Hamiltonian, 
$\phi_{\bf k}$ and $\phi_{\bf z} = (2\pi)^{-3}\int d^3 k e^{i{\bf k}\cdot{\bf z}}\phi_{\bf k}$ are the field amplitudes defined 
in the wave-vector space and in the position space, respectively, 
${\cal N} = \prod_{\bf k}(\omega_{\bf k}\epsilon/\hbar\pi)^{1/4}$ is the normalization constant with
the volume element $\epsilon \equiv 1/[(2\pi)^3 \delta^3(0)] \sim d^3 k/(2\pi)^3$, 
$E_0$ is the vacuum energy, $\omega_{\bf k} \equiv \sqrt{|{\bf k}|^2+m^2}$, and the function
\begin{equation}
 g({\bf z}-{\bf z}') \equiv \int {d^3k\over(2\pi)^3}\omega_{\bf k}e^{-i{\bf k}\cdot ({\bf z}-{\bf z}')}
\end{equation}
is real and nonvanishing in general for ${\bf z}\not= {\bf z}'$. 
The one-particle state of the field mode $\phi_{\bf p}$ reads \cite{LCH09}
\begin{eqnarray}
\Psi_{1_{\bf p}} [ \phi;t ] &=& \langle\phi |e^{-i\hat{H}t/\hbar}| 1_{\bf p}\rangle = {\cal N} e^{-i E_0 t/\hbar} \, 
  \hat{b}^\dagger_{\bf p} \, \exp -{1\over 2\hbar}\int {d^3 k\over (2\pi)^3}\omega_{\bf k} \phi_{\bf k}\phi_{\bf -k}
  \nonumber\\ &=& {\cal N} e^{-i (E_0 + \hbar\omega_{\bf p}) t/\hbar}
  \sqrt{2\omega_{\bf p}\epsilon\over \hbar}\phi_{-\bf p} 
  \exp -{1\over 2\hbar}\int {d^3 k\over (2\pi)^3}\omega_{\bf k} \phi_{\bf k}\phi_{\bf -k} \nonumber\\ &=& 
  {\cal N} e^{-i (E_0 + \hbar\omega_{\bf p}) t/\hbar} \sqrt{{2\omega_{\bf p}\epsilon\over \hbar}}
  \int d^3 x e^{i {\bf p}\cdot{\bf x}}\phi_{\bf x} \times \nonumber\\ & & \;\;\; \exp
  -{1\over 2\hbar}\int d^3 z d^3 z' \phi_{\bf z} g({\bf z}-{\bf z}') \phi_{{\bf z}'} .
\label{1particle}
\end{eqnarray}
Obviously neither $\Psi_0 [ \phi; t ]$  nor $\Psi_{1_{\bf p}} [ \phi;t ]$ 
is a product or separable state in terms of $\phi_{\bf z}$ ($\not\sim \prod_{\bf z} \psi(\phi_{\bf z})$),
while both of them are separable states in terms of $\phi_{\pm\bf k}$.

Consider an one-particle state of a wave packet $\varphi({\bf x}) = \int {d^3 p\over (2\pi)^3} e^{i {\bf p}\cdot{\bf x}} 
\tilde{\varphi}({\bf p})$ instead of a single field mode $e^{i {\bf p}\cdot{\bf x}}$ \cite{SR07, BBBD12}, 
the wave-functional reads
\begin{eqnarray}
& &\Psi_{1_{\varphi}} [ \phi;t ] \nonumber\\ 
&=& {\cal N} e^{-i E_0 t/\hbar}\left[ \int {d^3 p\over (2\pi)^3\sqrt{\epsilon}}  e^{-i \omega_{\bf p}t} 
    \tilde{\varphi}({\bf p})\hat{b}^\dagger_{\bf p}\right]
    \exp -{1\over 2\hbar}\int {d^3 k\over (2\pi)^3}\omega_{\bf k} \phi_{\bf k}\phi_{\bf -k} \nonumber\\ 
  &=& {\cal N}  \int {d^3 p\over (2\pi)^3} e^{-i(E_0 + \hbar \omega_{\bf p})t/\hbar} \tilde{\varphi}({\bf p})
    \sqrt{{2\omega_{\bf p}\over \hbar}} \phi_{-\bf p} 
    e^{ -{1\over 2\hbar}\int {d^3 k\over (2\pi)^3}\omega_{\bf k} \phi_{\bf k}\phi_{\bf -k}} 
\label{1particleWP}
\end{eqnarray}
where $1 = \int d^3 x |\varphi({\bf x})|^2 = \int {d^3 p\over (2\pi)^3} | \tilde{\varphi}({\bf p})|^2$ and
$[\hat{b}^{}_{\bf k}, \hat{b}_{\bf p}^\dagger] = \delta^3({\bf k}-{\bf p})/\delta^3(0)$ so that $\Psi_{1_{\varphi}}$ is normalized.
One can see that only the polynomial part (in the braced bracket above) of the wave-functional is modified from (\ref{Minkvac}) 
or (\ref{1particle}) 
and the exponent ($\sim \int {d^3 k\over (2\pi)^3}\omega_{\bf k} \phi_{\bf k}\phi_{\bf -k} = \int d^3 z d^3 z' 
\phi_{\bf z} g({\bf z}-{\bf z}') \phi_{{\bf z}'}$) is still the same.
Even if the wave packet $\varphi({\bf x})$ is extremely localized in space the field state will still be entangled
in terms of $\phi^{}_{\bf x}$ due to the nonlocality of $g({\bf z}-{\bf z}')$ in the exponent. \\ \\

\end{appendix}

\end{document}